\documentclass{iau} 
\usepackage{graphicx}

\title[Astronomy Education in Covid Times] 
{Astronomy Education in Covid Times}

\author[Priya Hasan \& S N Hasan ]{Priya Hasan \& S N Hasan}   

\affiliation{Maulana Azad National Urdu University, Hyderabad, India \\ email: {\tt priya.hasan@gmail.com} }

\pubyear{2021}
\volume{367}  
\jname{Education and Heritage in the era of Big Data in Astronomy}
\editors{R.M. Ros, B. Garcia, S. Gullberg, J. Moldon \& P. Rojo, eds.}

\begin{document}

\maketitle

\begin{abstract}
We shall describe the various activities done by us in Covid Times including outreach and educational workshops in Physics  and Astronomy. We shall discuss the caveats in virtual teaching of Astronomy and the lessons learnt in the process. 
\end{abstract}

\firstsection 
\section{Introduction}

The pandemic of 2020 came as a shock and a challenge to the teaching community all over the world. It meant adapting to a new system of communication as the sole source with its limits on interaction and reach. It was a learning experience for teachers as well as students because of its strong dependence on pedagogical methodologies and  technologies. 


\section{Activities}

 We collaborated with Jana Vigyan Vedika (JVV) Telangana which is a member organisation of the All Indian People Science Network (AIPSN) and  a member of the National Council for Science and Technology Communication (NCSTC)in India. The NCSTC is a scientific programme of the Government of India for the popularisation of science, dissemination of scientific knowledge and inculcation of scientific temper. JVV is an effective Science Forum in the state and has a very good reach in rural and urban  Telangana. 

We started off with a Six-Day Online Astronomy Class from the 18-23 May 2020 from 4:30- 6:30 pm IST. We conducted a quiz at the end of the event and gave certificates to participants who got the minimum mark. We had 475 registrations and many were given certificates with JVV.
 
Encouraged by the success of this, we conducted a Six-Day Workshop `Joy of Learning Physics@ Home' from 15-20 June 2020 for school students.
We showed them live demonstratons on Newton's Laws, Gravity, Simple Machines, Pressure, Optics and  Electricity \& Magnetism. At the end of each session, we demonstrated an experiment with questions to be answered by students the next day. It was very interactive and students were very enthusiastic about it. We had an International-Astronomical Union-Office of Astronomy for Development (IAU-OAD) Project `Clear Skies' for school children and hence alot of the material for activities were available to us. 
 
In the meanwhile, the IAU-OAD announced a special call for Covid Times Projects. We proposed a project `Astronomy from Archival Data' which involved Educational Activities for Under-Graduate  and Post-Graduate Students (\cite[Possel, 2020]{Possel20}). The project is close to completion. For the four months of August-November 2020, we had sessions every weekend on Saturdays and Sundays where students learnt python, astropy, virtual observatory tools like Topcat (\cite[Taylor, 2003]{Taylor93}), Aladin, ESASky and sources of archival data. Resource persons from all over the world contributed in guiding students in 10 Research Projects. The Internal Virtual Observatory Alliance (IVOA) supports this project. Participants will present their projects in the end of January. A more detailed report of this is presented in this series in the form of a talk. 

Prior to the pandemic, we had received funding from the US Consulate, Hyderabad to conduct a Hands-on Astronomy Workshop for School Teachers. Initially we were waiting for the situation to improve, but when it did not, we got 68 astronomy kits delivered to School Teachers all over the country. We then had a Two-Day Hands-on Astronomy Workshop online for almost 100 teahers on the 30-31 October 2020. This Workshop covered topics like: Day-time Astronomy activitiies, Sun-Dials, Optics, Telescope Malking and a basic introduction to Astronomy. We also had a Panel Discussion on application of Mathematics and Physics concepts in Astronomy.

To add on to the recent activity, we had another Hands-on Astronomy Workshop on the 2,9 and 16 December with sessions by Prof Daniel Barth on Low-cost Activities in Astronomy\footnote{Barth, D. E. (2019). Astronomy for Educators. Open Educational Resources. Available at: https://scholarworks.uark.edu/oer/2}.

As a team with Prof S N Hasan, we run Shristi Astronomy (https://shristiastro.com/). Details of our activities are on the website and the reader is encouraged to have  a look. All the sessions have been recorded and are available on youtube\footnote{https://www.youtube.com/c/shristiastronomy}. 
  
\section{Lessons Learnt}
In the course of our sessions we learned the following:
\begin{itemize}
\item
Planning of the events has to be done carefully since most potential participants had their own online sessions on. We selected weekends, evenings or holidays for all our events. 
\item There are various issues like internet connectivity problems, power as well as overlapping events. Hence all our activities were recorded and made available on youtube  so that participants could attend at their convenience. 

\item A few of the participants were from non-English speaking countries, hence we enabled subtitles on our videos for their benefit. 
\item
It is good to have a variety of speakers ranging from students, research scholars, post-doctoral fellows and junior and senior faculty. The level needs to be matched to the average participant.  
\end{itemize}
We hope that these resources we created would be useful even when better times return.

\end{document}